# Social Expansion versus Social Fragmentation

*Don't take Europe for granted*

Ingo Piepers
Tower Projects
Version: 30 September 2008

E-mail: ingopiepers@gmail.com

**Abstract**

The process of social expansion in Europe can be better understood with various concepts related to complexity science. Findings of exploratory research show a typical process of social expansion in Europe in the period 1495-1945, in which wars have been instrumental. Furthermore, this research enables the identification of vulnerabilities, and the conditions for success in a process of social expansion.

## 1. Introduction

European unity, the European Treaty, the limits of expansion of the European Union (EU), and how the EU should be governed – just to name some subjects – are extensively debated. The outcome of these discussions, and the discussions themselves, are of fundamental importance for the future of Europe. Will Europe be able to further consolidate the various forms of cooperation between its member states, or will – sooner or later – competition and conflict prevail, and as a consequence renationalization become unavoidable? Is Europe's future: consolidation and expansion or fragmentation?

Typically, the discussions about Europe's future consist of the exchange of qualitative arguments, and are often only superficial. Problematic is that historical facts – as far as they are known and properly understood – are often selectively interpreted. Local interests of decision makers and their inability to understand what is at stake, make this often unavoidable. These biases often hinder decision-makers in Brussels and the capitals of the EU's member states.

In order to speed up Europe's unification, these discussions should - if possible - be objectified. The unification of Europe is important – as I will explain in this paper – because Europe – as the rest of the world **-** will unavoidably be confronted with global and regional problems that require intense cooperation, in order to avoid 'devastating' consequences of these developments.

In this paper I will show that complexity science can contribute to a more objective approach of these European challenges. I will show that the development of Europe towards a social cohesive 'unit' can be quantified: Europe is not a virtual reality or an artificial entity, but a hard 'fact'.

In short, the following is made clear by this research: In the last 500 years, Europe has developed towards a more stable condition; wars were instrumental in this process of social expansion; and this process has a certain logic to it, as I will show. However, on the shorter term, the war dynamics between states were to a high degree unpredictable, as a consequence of – as I will argue – a chaotic attractor 'normally' influencing the war dynamics of the International System.

In the International System chaos and order go hand in hand, and are closely related properties of this system.

Furthermore this research not only shows that a number of properties of the International System can accurately be quantified, but that these properties follow a clear developmental path. A stable - conflict free - Europe is the outcome of this process of social expansion, however - as I will argue - a conflict free Europe is not automatically preserved: a number of conditions have to be met in order to avoid the resumption of the typical war dynamics of the European International System.

In this paper I will discuss these research findings in more detail, in an effort to objectify discussions about Europe's future, because new insights and concepts could well lead to new problem definitions and solutions. For a more technical discussion of the research findings, I refer to *Self-Organized Characteristics of the International System* (Piepers, 2007).

## 2. The International System as a Complex System

Complexity science is a relatively new scientific discipline, and focuses on systems that typically consist of a relatively large number of elements or actors, that interact on the basis of changing rules. Complexity science has its roots/fundaments in theoretical physics, and has contributed to new insights in the functioning of our climate and ecosystems.

These systems often show a certain degree of self-organization – emergence – whereby interactions between the actors of the system result in unforeseen/unplanned structures and regularities at system – that is – macro level. Typically these systems show non-linear behaviour and effects; in case of these interactions cause and effect are not proportional.

This research suggests that the International System could probably be characterised as a complex system. States are the actors of this system, which interact on the basis of changing rules, e.g. international law. These laws more or less continuously change, on the basis of the requirements of its member states – especially Great Powers – and the International System itself. Great Powers can accurately be indentified on the basis of various 'hard' characteristics (Levy, 1983).

Four categories of basic functions/requirements of social systems – states in this context – can be identified (Piepers, 2006). The requirement for: (1) Internal and external security, and the potential to influence the behaviour of individuals and other (sub) systems, (2) Energy, necessities of life, and wealth, (3) Individual and collective identity and the development of these identities, and (4) (a) Internal and external consistency of the system itself, (b) direction for the (future) development of the social system, (c) acceptance of the (political) leadership of the social system, and (d) the possibility to control the environment of the social system. The integrative function is responsible for the coordinated fulfilment of the (future) requirements of social systems.

These four requirements constitute subsystems, respectively: a threat system, the economy, the culture of the state, and the integrative system of the state. These subsystems have their own typical dynamics and they often compete for priority, however a minimum level of 'fulfilment' of all four needs is required in order to guarantee the survival of the social system (a state, the EU and the International System in this context).

From this perspective the function of the European Union and the European Treaty is to structure and manage this balancing act, and to ensure the simultaneous fulfilment of the basic requirements of the EU's member states, the EU itself, and preserve the integrity of the EU.

## 3. Properties of the International System, influencing its Dynamics and Development

Before elaborating on the findings of my research, it is important to discuss three properties of the International System.

First, the International System is an anarchistic system. At a global level the International System lacks an accepted 'overall' authority, which can determine (democratically or otherwise) the goals, priorities and organization of the International System. In an anarchistic system, states are responsible for their own security. This property results in a 'security dilemma'. In order to enhance their own security, states will arm themselves and participate in alliances, however one state its (improved) security is another state's insecurity (Holsti, 1995). Depending on various conditions of the International System (e.g. the polarity of the system), this typical property of an anarchistic system can result in a positive feedback mechanism; an arms race.

The life cycle of Great Powers, concerning their capacities and status (power) is the second property.

The (relative) power of states, and the resulting ranking in the International System, is not static, and shows a typical life cycle. In this life cycle various factors play a role: technology and economics, as well as (international) obligations, requirements for military investments, and the flexibility of the institutions of states. As a consequence of changes in these factors, the (relative) power positions of states change continuously (Gilpin, 1981).

Military power has - especially in case of 'overstretch' - always a downside. Military power does contribute to a state's power position and status, but the required investments sooner or later affect the potential for economic growth and development, especially if an arms race has become 'unavoidable'. Great Powers tend to become victims of their own success.

The life cycle of states – especially of Great Powers – contribute to the build up of tension and frustration in an anarchistic International System. 'Power dynamics' not only influence the position and ranking of states in the International System and the system's prestige hierarchy, but also have an impact of the threat perception of these states. A (feeling of) diminished power reinforces a feeling of vulnerability and makes decision-makers more insecure.

The third property concerns the functioning and the development of international law and its institutions. Since the 'forming' of the International System – in the late 15th century – the International System has some ordering. This ordering is embedded in the laws and institutions of the system, and has regulated the (inter)actions of states. These laws and institutions – especially their 'irregular' development – have had a profound impact on the dynamics and development path, of the International System as well.

The most important laws and institutions include the sovereignty principle (1648), the Concert of Europe (1815), the League of Nations (1919), and the United Nations (1945).

## 4. Systemic Wars and Periodic Reorganization of the International System

Above mentioned laws and institutions did not come about arbitrarily or coincidentally, but were always the outcome of large-scale Great Power wars - systemic wars – respectively: the Thirty Years War, the Napoleonic Wars, and the First and Second World War. In fact, these systemic wars constitute reorganizations of the anarchistic International System. Obviously, the anarchistic International System lacks other means than war to periodically rebalance the (power) relationships and interactions in the International System.

As is the case with Great Powers, the international system - more specifically the ordering of the system - has a typical life cycle and life span. International Systems become obsolete, and relatively stable periods are periodically punctuated by systemic wars; as explained: the purpose of these systemic wars is reorganization. On the longer term it is possible to discern a development path in the ordering of consecutive international systems.

It can be observed that in the longer term the acceptance of the use of war as a legitimate instrument in international relations has diminished. New laws have progressively delegitimized the use of violence, and have – as a consequence - resulted in certain thresholds in the International System.

These three properties of the International System – its anarchistic structure resulting in a security dilemma, the life cycle of Great Powers, and the development of the International System, progressively reducing the acceptance of the use of violence, to a high degree determine the functioning of the International System as a complex system.

## 5. SOC-Characteristics of the International System

In complexity science a special category of systems is distinguished: systems that 'organize' themselves into a critical condition. These systems are called SOC –systems; SOC stands for Self-Organized Criticality.

Typically, in these systems a 'driving force' is at work, resulting in the build-up of tension in the system. Thresholds in these systems allow for the build up of tension, preventing immediate relieve. Incidents periodically trigger release events. After such a relieve event the system has new 'space' for the build up of tension, and the process can repeat itself. The size and number of the release events in these SOC-systems show a remarkable statistic relationship: A power law.

The research results show that the International System has – at least in the period from 1495 until 1945 – SOC-characteristics.

Conflicts of interest between states, and (increasing) frustration over the functioning of the International System constitute the driving force of the International System, and result in the build up of tension. International laws and institutions form the thresholds of the International System. Interests of states have a threshold effect as well, e.g. the interests of Great Powers to maintain the status quo.

From this perspective wars between states can be considered the release events of the International System. The start of the First World War, the murder of Franz Ferdinand in1914 in Sarajevo, shows how a relatively minor incident can trigger a non-linear release event of systemic size; the First World War. Obviously the International System was at that stage in a critical condition.

It is in accordance with the typical characteristics of a SOC-system, that the size and number of Great Power wars in the period 1495-1945, can be shown with a power law: The number and size of wars are not arbitrary, but show clear regularities.

As discussed, four large-scale – systemic – wars can be identified. These four wars have resulted in the periodic reorganization - realignment - of the International System.

## 6. Exceptional War Dynamics: Providing Additional Insights in the Workings of the International System

During the period 1657-1763, a number of large-scale wars took place as well. However, these wars were of a different category, and did not result in the fundamental reorganization of the International System.

From a complexity perspective it can be argued, that during this specific time frame, the International System lacked 'degrees of freedom', restricting the typical chaotic dynamics of the system. As a consequence a simplified, more predictable "quasi periodic" war dynamic emerged.

This exceptional period can be interpreted from a historical perspective as well. Historians have noticed that during this time frame the international system functioned quite flexible. During this period the dynamics of the International System were to a high degree dominated by the intense rivalry between France and Great Britain. It seems that other Great Powers temporarily did not influence the war dynamics of the International System.

The Cold War had a similar 'ossifying' effect on the war dynamics and development of the International System, especially in Europe. The end of the Cold War resulted in an increase in the degrees of freedom in the International System, and causing a 'return' of the (typical) chaotic dynamics of the International System.

## 7. SOC and Punctuated Equilibrium Dynamics of the International System

In complexity science, the typical dynamic of the International System - the exceptional period not taken into account - during which relatively stable periods (with only 'minor' wars) were punctuated by large scale and intense systemic wars, is denoted as a punctuated equilibrium dynamic. Typically during a punctuation – as is the case with the International System as well – a more fundamental development of the system takes place: It is during systemic wars that new rule sets for the International System are formulated and embedded. As explained the (new) Great Powers emerging from the latest systemic war, to a high degree determine the outline of the new international system.

This research shows that the SOC-characteristics and the punctuated equilibrium dynamics of the International System are closely related.

## 8. Development of Europe towards a Stable - Conflict Free - Condition

In case a number of quantitative characteristics of the relatively stable periods - the periods between punctuations/systemic wars - of the International System are more closely researched, it shows that some interesting trends can be discerned in the development of the International System.

For example it becomes clear that during the stable periods (1495-1618, 1648-1792, 1815-1914, 1918-1939) the number of Great Power wars linearly declined.

If this analysis is restricted to wars of European Great Powers (in the period 1495-1945, just a very small fraction of the total number of Great Power wars were *not* European Great Power wars), than – in line with this linear decrease – the number of Great Power wars after the Second World War in Europe should be reduced to (almost) nil. And this is indeed the case.

A conclusion of this research is not only that (systemic) wars are instrumental in the periodic reorganization of the International System, but functional in a (European) process of social expansion as well. By the process of social expansion in Europe, I point to the process in which various forms of cooperation were/are formed between states in Europe. This process includes the institutionalization of rules and forms of governance. During this process, which started in 1495, but is still not finalized, Europe has become stable, and conflict has been replaced by cooperation.

**9. The Mechanisms and Conditions resulting in Stability in Europe**

A relevant question is what mechanism resulted in the linear stabilizing of Europe and the concurrent process of social expansion. Complexity science can provide some suggestions.
Simulations with simple models of networks show a relationship between the size and frequency of so called cascades that occur in these systems on the one hand, and the connectivity and thresholds of these systems on the other hand (Watts, 2002).
Cascades are in fact local reorganizations of the system, comparable to release events. With the connectivity of the system is meant the number of connections of a node (an actor) with other nodes (actors). The threshold of the actors of the system determines when a node (actor) switches of condition (e.g. opinion). The fraction of the number of connections of an actor that are required to change (of position or opinion for example), before the actor itself changes of position or opinion, determines the threshold.
These simulations show that by certain combinations of the connectivity of the system and thresholds, cascades become impossible. High connectivity and high thresholds result in a high local stability of the system, making cascades impossible: the system has become stable.
These simulations provide some clues for a better understanding of the stabilizing process and process of social expansion in Europe.
Wars can be perceived as cascades in the International System. Furthermore, it can be determined that during the period under consideration (1495-1945) the connectivity of the International System- and certainly of Europe – has steadily increased. This for example is valid for the number, reach and intensity of economic interactions between states, and the 'exchange' of norms and values between actors of the International System. In addition, the thresholds of the International System were periodically enhanced. As discussed, the punctuations - systemic wars – were instrumental in this development of the thresholds of the International System. The development of these two characteristics of the International System probably explains the (linear) decrease of the number of Great Power wars in Europe.
Ultimately – that is the conclusion based on these simulations – as a consequence of the increased connectivity and thresholds in Europe, the security dilemma and the thus the driving force were neutralized.
But this is only part of the explanation for the process of social expansion. Other factors were – are – important as well.
Another important factor in the process of social expansion in Europe has been the gradual development of a shared perspective on the future of Europe, facilitated and stimulated by the troubled history of the continent.
The Cold War has played on important role as well. The Cold War provided favourable conditions for the structural neutralization of the security dilemma in Europe. The Cold War - with the United States and the Soviet Union as its two major protagonists - deprived (European) states of the 'opportunity' to wage war: the system was 'frozen', severely restricting the war dynamics in the International System. The Cold War thus provided an opportunity to intensify cooperation; economic, political and military, and time to embed – institutionalize – this cooperation.
It is important that in this new European System, the release of tension and frustration can be facilitated, without the 'necessity' of going to war. Further institutionalization of the connectivity of the system (political and economic cooperation), thresholds concerning the use of conflict, and the development of a shared perspective on Europe's future, are essential building blocks of a stable and prosperous Europe as well. But even this is not enough.

An effective political system and process to regulate tensions and frustration are required as well. Democracy is such a political system. The evolutionary success of democracy as an effective political system is convincing evidence: democracy and survival are closely related, as history shows.

## 10. A Less Attractive Alternative: Fragmentation

The following process could evolve if the process of social expansion is derailed, that is if proper consolidation is not achieved: cooperation starts to falter, and will be replaced by competition, the connectivity of the system will be affected as a consequence, democratic structures – as far as they are now available – will become (even) less effective, hindering the release of tension and frustration. National interests will replace common – European – interests and next, the local stability of the system will be affected, and the security dilemma reactivated, 'enabling' conflicts. In this scenario the (European) process of social expansion is replaced by a process of social fragmentation.

Fragmentation processes are not new to our International System; failed states are a typical example. So don't take stability in Europe for granted. Stability needs constant effort and commitment to be maintained.

## 11. Urgency is required: The Centrifugal Effects of Global Challenges and Threats

Urgency is required to further develop and embed institutional structures in Europe. Europe, as will the rest of the world, will be confronted with a series of challenges and threats, that require more than just a local - that is state-level or regional - response: climate change, (nuclear) terrorism, poverty, shortage of commodities, etc. etc. These challenges and threats will test Europe's fragile fabric, and the willingness of its member states to cooperate and compromise.

These research findings not only provide insight in the functioning and in the conditions for success for an effective European Union, but can contribute to the Europe debate as well. We now better understand what the EU's vulnerabilities are, what our priorities should be and what we should be talking about